\documentstyle[pic,epsfig]{article}

\let\citep\cite

\newcommand{\bq}{\begin{quotation}}
\newcommand{\eq}{\end{quotation}}

\begin{document}

\title{Quantum Bayesianism at the Perimeter}
\author{Christopher A. Fuchs}
\maketitle

\section{Abstract}

{\small
The author summarizes the Quantum Bayesian~\cite{Fuchs10b,Caves02,Fuchs02,Fuchs04,Caves07,Fuchs09a,Fuchs10} viewpoint of quantum mechanics, developed originally by C. M. Caves, R. Schack, and himself.  It is a view crucially dependent upon the tools of quantum information theory.  Work at Perimeter Institute continues the development and is focussed on the hard technical problem of a finding a {\it good\/} representation of quantum mechanics purely in terms of probabilities, without amplitudes or Hilbert-space operators.  The best candidate representation involves a mysterious entity called a symmetric informationally complete quantum measurement.  Contemplation of it gives a way of thinking of the Born Rule as an {\it addition\/} to the {\it rules\/} of probability theory, applicable when one gambles on the consequences of interactions with physical systems.  The article ends by outlining some directions for future work.\footnote{This article is an extract from a longer one titled, ``\ldots And the Perimeter of Quantum Bayesianism.''  Beside expanding on all the points here, it gives a Quantum Bayesian answer to the {\it crucial\/} question ``If quantum states are not part of the stuff of the world, then what is?''  See Ref.~\cite{Fuchs10b}.  }
}

\section{A Feared Disease}

In these days of unceasing media coverage for the H1N1 pandemic, we are reminded that a healthy body can be stricken with a fatal disease that to outward appearances is nearly identical to a common yearly annoyance: simple seasonal flu.  There are lessons here for quantum mechanics.  In the history of physics, there has never been a healthier body than quantum theory; no theory has ever been more all-encompassing or more powerful.  Its calculations are relevant at every scale of physical experience, from subnuclear particles, to table-top lasers, to the cores of neutron stars and even the first three minutes of the universe.  Yet, since its founding days, many physicists have feared that quantum theory's continuing annoyance---the unease that something at the bottom of the theory does not make sense---may one day turn fatal itself.

There is something about quantum theory that is different in character from any physical theory posed before.  To put a finger on it, the issue is this:  The basic statement of the theory---the one we have all learned from our textbooks---seems to rely on terms our intuitions balk at as having any place in a fundamental description of reality.  The notions of ``observer'' and ``measurement'' are taken as primitive, the very starting point of the theory.  This is an unsettling situation!  Shouldn't physics be talking about {\it what is\/} before it starts talking about {\it what will be seen\/} and who will see it?  Perhaps no one has put the point more forcefully than John Stewart Bell~\cite{Bell90}:
\begin{quote}\small
What exactly qualifies some physical systems to play the role of `measurer'?  Was the wavefunction of the world waiting to jump for thousands of millions of years until a single-celled living creature appeared?  Or did it have to wait a little longer, for some better qualified system \ldots\ with a PhD?
\end{quote}
One sometimes gets the feeling that until this issue is settled, fundamental physical theory has no right to move on.  Worse yet, that to the extent it does move on, it does so only as the carrier of something insidious, something that will eventually cause the whole organism to stop in its tracks.  ``Dark matter and dark energy?  Might these be the first symptoms of a something systemic?  Might the problem be much deeper than getting our quantum fields wrong?'' --- This is the kind of fear at work here.

So the field of quantum foundations is not unfounded; it is absolutely vital to physics as a whole.  But what constitutes ``progress'' in quantum foundations?  Throughout the years, it seems the most popular criterion has derived from the tenor of Bell's quote:  One should remove the observer from the theory just as quickly as possible.  In practice this has generally meant to keep the {\it mathematical structure\/} of quantum theory as it stands (complex Hilbert spaces, etc.), but find a way to tell a story about the mathematical symbols that involves no observers.

Three examples suffice to give a feel:  In the de~Broglie~--~Bohm ``pilot wave'' version of quantum theory, there are no fundamental measurements, only ``particles'' flying around in a $3N$-dimensional configuration space, pushed around by a wave function regarded as a physical field.  In ``spontaneous collapse'' versions, systems are endowed with quantum states that generally evolve unitarily, but from time-to-time collapse without any need for measurement.  In Everettian or ``many-worlds'' quantum mechanics, it is only the world as a whole---they call it a multiverse---that is really endowed with an intrinsic quantum state. That quantum state evolves deterministically, with only an {\it illusion from the inside\/} of probabilistic ``branching.''

The trouble with all these interpretations as quick fixes to Bell's complaint is that they look to be just that, {\it really quick fixes}.  They look to be interpretive strategies hardly compelled by the details of the quantum formalism.  This explains in part why we could exhibit three such different strategies, but it is worse:  Each of these strategies gives rise to its own set of incredibilities---ones for which, if one were endowed with Bell's gift for the pen, one could make look just as silly.  Take the pilot-wave theories:  They give instantaneous action at a distance, but not actions that can be harnessed to send detectable signals.  If there were no equations pretending a veneer of science, this would have been called counting angels on the head of a pin.

\section{Quantum States Do Not Exist}

There is another lesson from the H1N1 virus.  To some perplexity, it seems people over 65---a population usually more susceptible to fatalities with seasonal flu---fare better than younger folk with H1N1.  No one knows exactly why, but the leading theory is that the older population, in its years of other exposures, has developed various latent antibodies.  The antibodies are not perfect, but they are a start.  And so it may be for quantum foundations.

Here, the latent antibody is the concept of {\it information}, and the perfected vaccine, we believe, will arise in part from the theory of single-case, personal probabilities---the branch of probability theory called Bayesianism.  Symbolically, the older population corresponds to some of the founders of quantum theory (Heisenberg, Pauli, Einstein) and some of the younger disciples of the Copenhagen school (Rudolf Peierls, John Wheeler, Asher Peres), who, though they disagreed on many details, were unified on one point:  That quantum states are not something out there, in the external world, but instead are expressions of information.  Before there were people using quantum {\it theory\/} as a branch of physics there were no quantum states.  The world may be full of stuff, composed of all kinds of things, but among all the stuff and things, there is no observer-independent, {\it quantum-state kind of stuff}.

The immediate payoff of this strategy is that it eliminates the conundrums arising in the various objectified-state interpretations.  James Hartle~\cite{Hartle68} put the point decisively, ``The `reduction of the wave packet' does take place in the consciousness of the observer, not because of any unique physical process which takes place there, but only because the state is a construct of the observer and not an objective property of the physical system.''  The real substance of Bell's fear is just that, the fear itself.  To succumb to it is to block the way to understanding the theory.  Moreover, the shriller notes of Bell's rhetoric are the least of the worries:  The universe didn't have to wait billions of years to collapse its first wave function---wave functions are not part of the observer-independent world.

But this much of the solution is an elderly and somewhat ineffective antibody.  Its presence is mostly a call for more research.  Luckily the days for this are ripe, and it has much to do with the development of the field of quantum information---that  multidisciplinary field that includes quantum cryptography and quantum computation.  Terminology can say it all:  A practitioner in that field is just as likely to call any $|\psi\rangle$ ``quantum information'' as ``a quantum state.''  ``What does quantum teleportation do?''  ``It transfers {\it quantum information\/} from Alice to Bob.''  What we have here is a change of mindset~\cite{Fuchs10}.

What the protocols and theorems of quantum information pound home is the idea that quantum states look and feel like information in the technical sense of the word.  There is no more beautiful demonstration of this than Robert Spekkens's ``toy model'' mimicking various features of quantum mechanics \cite{Spekkens07}.  In this model, the ``toys'' are each equipped with four possible mechanical configurations; but the players, the manipulators of the toys, are consistently impeded from having more than one bit of information about each toy's actual configuration (two bits about two toys, etc.).  The only things the players can know are their states of uncertainty.  The wonderful thing is that these states of uncertainty exhibit many of the characteristics of quantum information:  from the no-cloning theorem to analogues of quantum teleportation, quantum key distribution, and even interference in a Mach-Zehnder interferometer.  More than two dozen quantum phenomena are reproduced {\it qualitatively}, and all the while one can pinpoint the cause:  The phenomena arise in the uncertainties, not in the mechanical configurations.

What considerations like this tell the objectifiers of quantum states is that, far from being an appendage cheaply tacked on to the theory, the idea of quantum states as information has a unifying power that goes a significant way toward explaining why the theory has the mathematical structure it does.  There are, however, aspects of Bell's challenge that remain a worry.  And upon these, all could still topple. Particularly, the questions {\it Whose information?}\ and {\it Information about what?}\ must be addressed before any vaccine can be declared a success.

Good immunology does not come easily.  But this much is sure:  The glaringly obvious (that a large part of quantum theory is about information) should not be abandoned rashly:  To do so is to lose grip of the theory, with no better grasp on reality in return.  If on the other hand, one holds fast to the central point about information, initially frightening though it may be, one may still be able to construct a picture of reality from the perimeter of vision.

\section{Quantum Bayesianism}

Every area of human endeavor has its bold extremes.  Ones that say, ``If this is going to be done right, we must go this far.  Nothing less will do.''  In probability theory, the bold extreme is personalist Bayesianism \cite{Bernardo94}.  It says that probability theory is of the character of formal logic---a set of criteria for testing consistency.  The key similarity is that formal logic does not have within it the power to set the truth values of the propositions it manipulates.  It can only show whether various truth values are inconsistent; the actual values come from another source.  Whenever logic reveals a set of truth values inconsistent, one must return to their source to alleviate the discord.  Precisely in which way to alleviate it, though, logic gives no guidance.

The key idea of personalist Bayesian probability theory is that it too is a calculus of consistency (or ``coherence'' as the practitioners call it), but this time for one's decision-making degrees of belief.  Probability theory can only show whether various degrees of belief are inconsistent. The actual beliefs come from another source, and there is nowhere to pin their responsibility but on the agent who holds them. A probability {\it assignment\/} is a tool an agent uses to make gambles and decisions, but probability {\it theory\/} as a whole is not about a single isolated belief---rather it is about a whole mesh of them.  When a belief in the mesh is found to be incoherent with the others, the theory flags the inconsistency.  However, it gives no guidance for how to mend any incoherences it finds.  To alleviate discord, one must return to the source of the assignments in the first place---the very agent who is attempting to sum up all his history and experience with those assignments.

Where personalist Bayesianism breaks from other developments of probability theory is that it says there are no {\it external\/} criteria for declaring an isolated probability assignment right or wrong.  The only basis for a judgment of adequacy comes from the {\it inside}, from the greater mesh of beliefs the agent accesses when appraising his coherence.  Similarly for quantum mechanics.

The defining feature of Quantum Bayesianism~\cite{Caves02,Fuchs02,Fuchs04,Caves07,Fuchs09a,Fuchs10} is that it says, ``If this is going to be done right, we must go this far.''  Specifically, there can be no such thing as a right and true quantum state, if such is thought of as defined by criteria {\it external\/} to the agent making the assignment:  Quantum states must instead be like personalist Bayesian probabilities.  The connection between the two foundational issues is this.  Quantum states, through the Born Rule, can be used to calculate probabilities.  On the other hand, if one assigns probabilities for the outcomes of a well-selected set of measurements, then this is mathematically equivalent to making the quantum-state assignment itself.  Thus, if probabilities are personal in the Bayesian sense, then so too must be quantum states.

What this buys interpretatively is that it gives each quantum state a home.  Indeed, a home localized in space and time---namely, the physical site of the agent who assigns it!  By this method, one expels once and for all the fear that quantum mechanics leads to ``spooky action at a distance,'' and expels as well any hint of a problem with ``Wigner's friend.''  It does this because it removes the very last trace of confusion over whether quantum states might still be objective, agent-independent, physical properties.

The innovation of Quantum Bayesianism is that, for most of the history of trying to take an informational point of view about quantum states, the supporters of the idea have tried to have it both ways:  that on the one hand quantum states are not real physical properties, yet on the other there is a right quantum state after all. One hears things like, ``The right quantum state is the one the agent should adopt if he had all the information.''  The tension in this statement, however, leaves its holder open to immediate attack:  ``If there's a right quantum state after all, then why not just be done with all this squabbling and call it a physical fact independent of the agent?  And if it is a physical fact, what recourse does one have for declaring that there is no action at a distance when delocalized quantum states change instantaneously?''

The Quantum Bayesian dispels these difficulties by being conscientiously forthright.  {\it Whose information?}  ``Mine!''  {\it Information about what?\/}  ``The consequences (for {\it me\/}) of {\it my\/} actions upon the physical system!''  The point of view here is that a quantum measurement is nothing other than a well-placed kick upon a system---a kick that leads to unpredictable consequences for the very agent who did the kicking.  What of quantum {\it theory}?  It is a {\it universal\/} single-user theory in much the same way that Bayesian probability theory is.  It is a users' manual that {\it any\/} agent can pick up and use to help make wise decisions in this world of inherent uncertainty.  In my case, a world in which $I$ am forced to be uncertain about the consequences of {\it my\/} actions; in your case, a world in which {\it you\/} are forced to be uncertain about the consequences of {\it your\/} actions.  In a quantum mechanics with the understanding that each instance of its use is strictly single-user---``My measurement outcomes happen right here, to me, and I am talking about my uncertainty of them.''---there is no room for most of the perennial quantum mysteries.

\begin{figure}
\begin{center}
\includegraphics[height=2.6in]{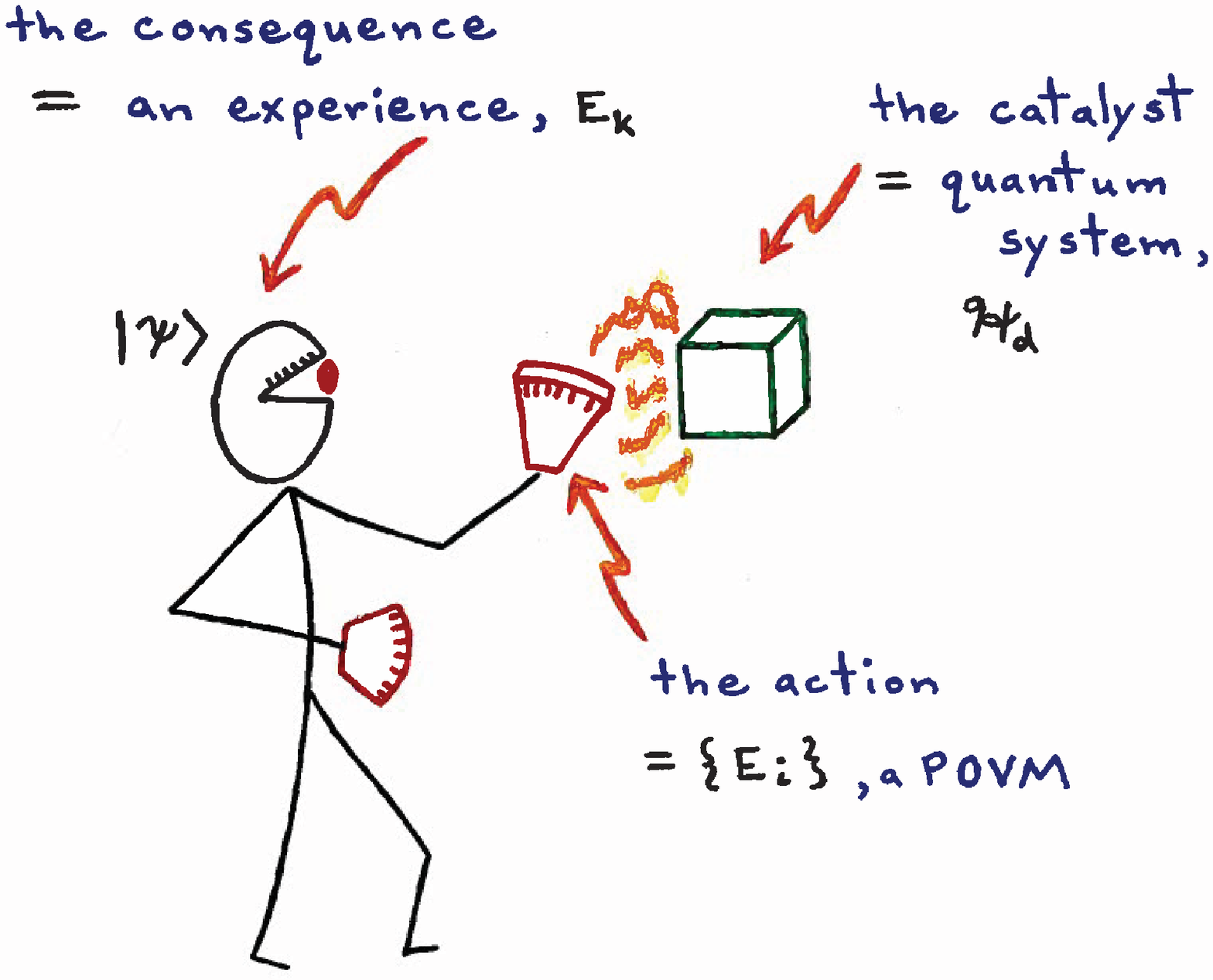}
\caption{}
\end{center}
\end{figure}

With this we finally pin down the way in which quantum theory is ``different in character from any physical theory before.''
For the Quantum Bayesian, quantum theory is not something {\it outside\/} probability theory---it is not a picture of the world as it is---but rather an {\it addition\/} to probability theory itself.  As probability theory is a {\it normative\/} theory, not saying what one {\it must\/} believe, but offering rules of consistency an agent should strive to satisfy within his overall mesh of beliefs, so it is the case with quantum theory.  To embrace this is all the vaccination quantum theory needs.

\section{Seeking SICs -- The Born Rule as Fundamental}

Yet, if quantum theory is a user's manual, one cannot forget that the world is its author.  And from its writing style, one may still be able to tell something of the author herself.  The question is how to tease out the motif.

Something that cannot be said of the Quantum Bayesian program is that it has not had to earn its keep in the larger world of quantum interpretations.  Since the beginning, the promoters of the view have been on the run proving technical theorems whenever required to close a gap in its logic or negate an awkwardness in its new way of speaking.  A case in point is the {\it quantum de Finetti theorem} \cite{Fuchs04,Caves02b}.

This is a theorem that arose from contemplating the meaning of one of the most common phrases of quantum information science---the unknown quantum state.  The term is ubiquitous:  Unknown quantum states are teleported, protected with quantum error correcting codes,  and much more.  From a Quantum-Bayesian point of view, however, it can only be an oxymoron:  If quantum states are states of belief, and not states of nature, then the state is known to someone, the agent who holds it.  But if so, then what on earth are experimentalists doing when they say they are performing quantum-state tomography in the laboratory?

The quantum de Finneti theorem is a technical result that allows the story of quantum-state tomography to be told purely in terms of a single agent---namely, the experimentalist in the laboratory.  In a nutshell, the theorem is this.  Suppose the experimentalist walks into the laboratory with the minimal belief that, of the systems his device is spitting out, he could interchange any two of them and it would not change the statistics he expects for any measurements.  Then the theorem says ``coherence alone'' requires him to make a quantum state assignment $\rho^{(n)}$ (for any $n$ of those systems) representable in the form:
\begin{equation}
\rho^{(n)}=\int P(\rho)\, \rho^{\otimes n} d\rho\;,
\label{MushuPork}
\end{equation}
where $P(\rho)\, d\rho$ is some probability measure on the space of single-system density operators and $\rho^{\otimes n}=\rho\otimes\cdots\otimes\rho$. To put it in words, this theorem licenses the experimenter to act {\it as if\/} each individual system has some state $\rho$ unknown to him, with a probability density $P(\rho)$ representing his ignorance of which state is the true one.  But it is only {\it as if}---the only real quantum state in the picture is the one the experimenter actually possesses, namely $\rho^{(n)}$.

This example is one of several, and what they all show is that the point of view has some technical crunch---it is not stale, lifeless philosophy.  Heartened with success, let us push toward a deeper question: If quantum theory is so closely allied with probability theory, why is it not written in a language that {\it starts\/} with probability, rather than a language that ends with it?  Why does quantum theory invoke the mathematical apparatus of Hilbert spaces and linear operators, rather than probabilities outright?  This brings us to present-day research at Perimeter Institute.

The answer we seek hinges on a hypothetical structure called a ``symmetric informationally complete positive-operator-valued measure,'' or SIC for short.  This is a set of $d^2$ rank-one projection operators $\Pi_i=|\psi_i\rangle\langle\psi_i|$ on a $d$-dimensional Hilbert space such that
\begin{equation}
\big|\langle\psi_i|\psi_j\rangle\big|^2=\frac{1}{d+1}\quad \mbox{whenever} \quad i\ne j\;.
\label{Mojo}
\end{equation}
Because of their extreme symmetry, it turns out that such sets of operators, when they exist, have remarkable properties.  Among these, two powerful ones are that they must be linearly independent (spanning the space of Hermitian operators) and sum to $d$ times the identity.

This is significant because it implies that an arbitrary state $\rho$ can be expressed as a linear combination of the $\Pi_i$.  Moreover, because the operators $H_i=\frac1d \Pi_i$ are positive-semidefinite and sum to the identity, these can be interpreted as labeling the outcomes of a quantum measurement device---not a von Neumann measurement device, but a measurement device of the most general kind allowed in quantum theory
\cite{Nielsen00}.  Finally, the $\Pi_i$'s symmetry gives a simple relation between the probabilities $P(H_i)={\rm tr}\big(\rho H_i\big)$ and the expansion coefficients for $\rho$:
\begin{equation}
\rho = \sum_{i=1}^{d^2}\left( (d+1)\,P(H_i) - \frac1d \right)\Pi_i\;.
\label{Ralph}
\end{equation}
The extreme simplicity of this formula suggests it is the best place for the Quantum Bayesian to seek his motif.

Before proceeding, we must reveal what is so consternating about the SICs: It is whether they exist at all.  Despite 10 years of growing effort since the definition was introduced \cite{Zauner99,Caves99}, no one has been able to show that they exist in general dimension.  All that is known firmly is that they exist in dimensions 2 through 67:  dimensions $2\,$--$\,15$, 19, 24, 35, and 48 by analytic proof, and the remainder through numerical simulation \cite{Scott09}.  How much evidence is this that SICs exist?  The reader must answer this for himself, but for the remainder of the article we will proceed as if they do for all finite dimensions $d$ and see where it leads.

\begin{figure}
\begin{center}
\includegraphics[height=3.3in]{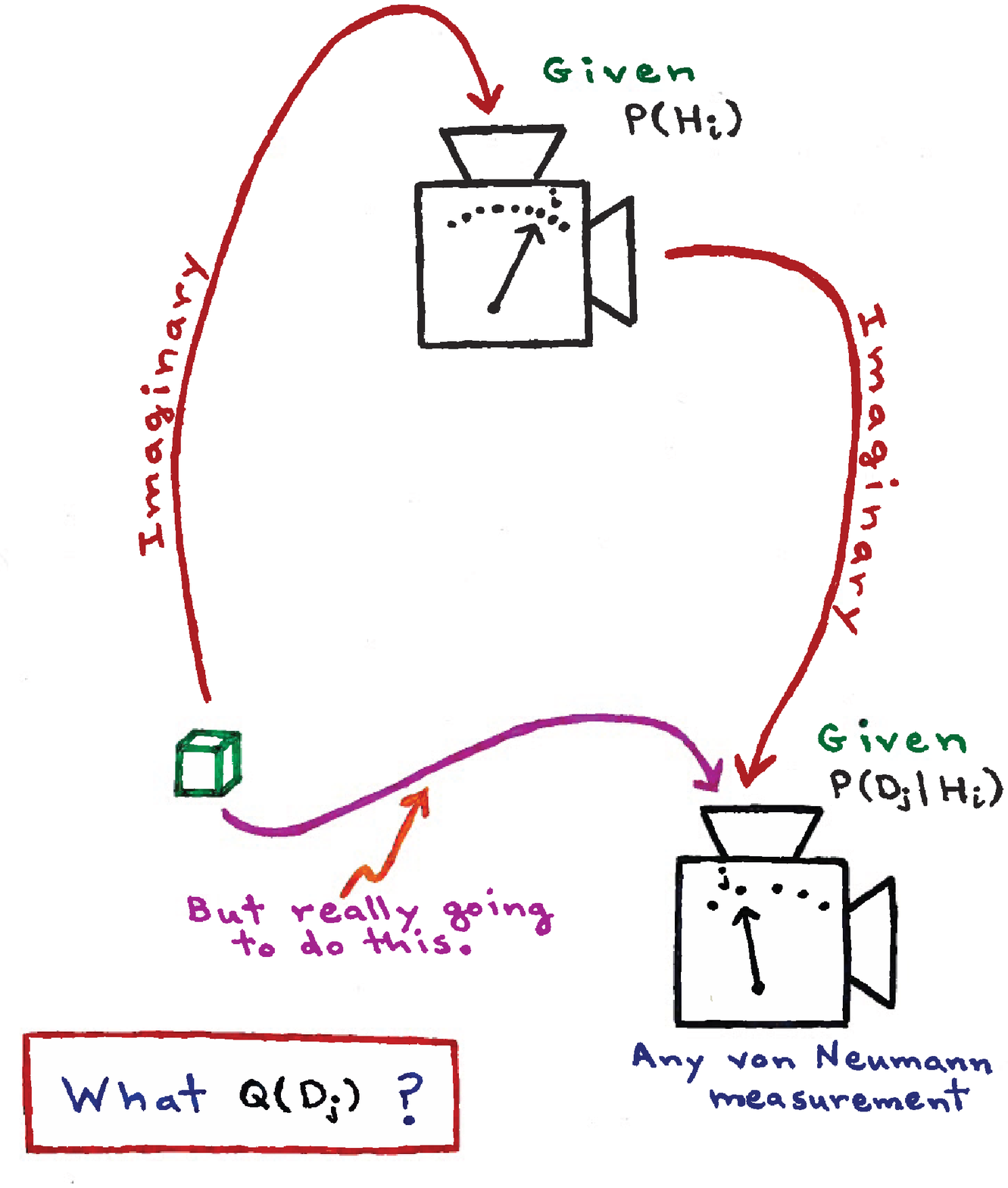}
\caption{}
\end{center}
\end{figure}

Thinking of a quantum state as {\it literally\/} an agent's probability assignment for the outcomes of a potential SIC measurement leads to a new way of expressing the Born Rule for {\it all\/} quantum probabilities.  Consider the diagram in Figure 2.  It depicts a SIC measurement ``in the sky,'' with outcomes $H_i$, and an arbitrary von Neumann measurement ``on the ground,'' with outcomes $D_j=|j\rangle\langle j|$, for some orthonormal basis.  We conceive of two possibilities (or two ``paths'') for a given quantum system to get to the measurement on the ground:  ``Path 1'' is that it proceeds directly to the measurement.  ``Path 2'' is that it proceeds first to the measurement in the sky and only subsequently cascades to the measurement on the ground.

Suppose now, we are given the agent's personal probabilities $P(H_i)$ for the outcomes in the sky and his conditional probabilities $P(D_j|H_i)$ for the outcomes on the ground subsequent to the sky.  I.e., we are given what the agent would assign on the supposition that the system follows Path 2.  Then ``coherence alone'' (in the Bayesian sense) is enough to tell what probabilities $P(D_j)$ the agent should assign for the outcomes of the measurement on the ground---it is given by the Law of Total Probability:
\begin{equation}
P(D_j)=\sum_i P(H_i) P(D_j|H_i)\;.
\label{Magnus}
\end{equation}
That takes care of Path 2, but what of Path 1?  Is this enough to recover the probability $Q(D_j)$ the agent would assign for the outcomes of Path 1 by the Born Rule?  That is, that
$
Q(D_j)={\rm tr} (\rho D_j)
$
for some quantum state $\rho$?  Clearly $Q(D_j)\ne P(D_j),$
for Path 2 is {\it not\/} a coherent process (in the quantum sense!)\ with respect to Path 1.

What is remarkable about the SIC representation is that it implies that, though $Q(D_j)$ is not equal to $P(D_j)$, it is nonetheless a function of it.  Particularly,
\begin{eqnarray}
Q(D_j) =
(d+1)\sum_{i=1}^{d^2} P(H_i) P(D_j|H_i) - 1\;.
\label{ScoobyDoo}
\end{eqnarray}
The Born Rule is nothing but a kind of Quantum Law of Total Probability!  No complex amplitudes, no operators---only probabilities in, and probabilities out.

Nonetheless, Eq.~(\ref{ScoobyDoo}) does not invalidate probability theory:  For the old Law of Total Probability has no jurisdiction in the setting of our diagram, which compares a ``factual'' experiment (Path 1) to a ``counterfactual'' one (Path 2).  Indeed as any Bayesian would emphasize, if there is a distinguishing mark in one's considerations, then one ought to take that into account in one's probability assignments.  Thus there is a suppressed condition in our notation:  Really we should have been writing the more cumbersome $P(H_i|{\mathcal E}_2)$, $P(D_j|H_i,{\mathcal E}_2)$, and $Q(D_j|{\mathcal E}_1)$ all along.  With this explicit, it is no surprise that $Q(D_j|{\mathcal E}_1)\ne P(D_j|{\mathcal E}_2)$. The message is that quantum theory supplies something that raw probability theory does not.  The Born Rule in these lights is an addition to Bayesian probability in the sense of giving an extra normative rule to guide the agent's behavior whenever he interacts with the physical world.

\section{The Future}

A vaccination of Quantum Bayesianism makes a healthy body even healthier, but it is far from the last word on quantum theory.  In fact it is just an indication of the great adventure that lies ahead.  By rewriting the Born Rule as Eq.~(\ref{ScoobyDoo}) one gets a sense of where the essence of quantum theory has been hiding all along.  It is in the active power of this quantity called dimension \cite{Fuchs10b}.  When an agent interacts with a quantum system, its dimension determines the extent to which the agent should deviate from the Law of Total Probability when transforming his counterfactual probability calculations to factual ones.  That ``power'' calls out for an independent characterization that makes no necessary reference to the agent using it.  Can it be done?  And if it can be done, what are its implications for physics as whole, from common laboratory issues to open questions in gravity and cosmology?  The Quantum Bayesians at Perimeter Institute are trying their best to find out.

\section{Acknowledgments}

The author thanks H.~C. von Baeyer for improving the presentation and M.~Schlosshauer for encouragement.  The ``Seeking SICs'' section was supported in part by the U.~S. Office of Naval Research (Grant No.\ N00014-09-1-0247).


\begin{thebibliography}{99}

\bibitem{Fuchs10b}
C.~A. Fuchs, {\tt arXiv:1001.????v1}

\bibitem{Caves02}
C.~M. Caves, C.~A. Fuchs and R.~Schack, Phys.\ Rev. A {\bf 65}, 022305 (2002).

\bibitem{Fuchs02}
C.~A. Fuchs, {\tt arXiv:quant-ph/0205039v1}.

\bibitem{Fuchs04}
C.~A. Fuchs and R.~Schack, {\tt arXiv:quant-ph/ 0404156v1}.

\bibitem{Caves07}
C.~M. Caves, C.~A. Fuchs, and R.~Schack, Stud.\ Hist.\ Phil.\ Mod.\ Phys.\ {\bf 38}, 255 (2007).

\bibitem{Fuchs09a}
C.~A. Fuchs and R.~Schack, {\tt arXiv:0906.2187v1}.

\bibitem{Fuchs10}
C.~A. Fuchs, {\sl Coming of Age with Quantum Information}, (Cambridge U. Press, 2010).

\bibitem{Bell90}
J.~S. Bell, Phys.\ World {\bf 3}, 33 (1990).

\bibitem{Hartle68}
J. B. Hartle, Am.\ J. Phys.\ {\bf 36}, 704 (1968).

\bibitem{Spekkens07}
R. W. Spekkens, Phys.\ Rev.\ A {\bf 75}, 032110 (2007).

\bibitem{Bernardo94}
J.~M. Bernardo and A.~F.~M. Smith, {\sl Bayesian Theory}, (Wiley, Chichester, 1994).

\bibitem{Caves02b}
C.~M. Caves, C.~A. Fuchs and R.~Schack, J. Math.\ Phys.\ {\bf 43}, 4537 (2002).

\bibitem{Nielsen00}
M. A. Nielsen and I. L. Chuang, {\sl Quantum Computation and Quantum Information}, (Cambridge U. Press, 2000).

\bibitem{Zauner99}
G. Zauner, PhD thesis, University of Vienna (1999).

\bibitem{Caves99}
C. M. Caves, report, University of New Mexico (1999).

\bibitem{Scott09}
A. J. Scott and M. Grassl, {\tt arXiv:0910.5784v1}.
\end{thebibliography}
\end{document}